# A Multimodal Assistive System for Helping Visually Impaired in Social Interactions

M. Saquib Sarfraz, Angela Constantinescu, Melanie Zuzej, Rainer Stiefelhagen

Karlsruhe Institute of Technology

Karlsruhe, Germany

## ABSTRACT

Access to non-verbal cues in social interactions is vital for people with visual impairment. It has been shown that non-verbal cues such as eye contact, number of people, their names and positions are helpful for individuals who are blind. While there is an increasing interest in developing systems to provide these cues less emphasis has been put in evaluating its impact on the visually impaired users. In this paper, we provide this analysis by conducting a user study with 12 visually impaired participants in a typical social interaction setting. We design a real time multi-modal system that provides such non-verbal cues via audio and haptic interfaces. The study shows that such systems are generally perceived as useful in social interaction and brings forward some concerns that are not being addressed in its usability aspects. The study provides important insight about developing such technology for this significant part of society.

## 1. INTRODUCTION

Social interactions are meetings or events that take place in a closed space, for instance in an office. People with vision impairments are at a disadvantage in social interactions as a face-to-face communication relies heavily on non-verbal cues. The access to the information *Who is looking at me*, for example, may avoid the awkward situation of answering a question that was directed to another person. This lack of access to non-verbal cues and the lack of awareness of people present in a social interaction may have adverse effects on the self-confidence of the people who have visual impairments and can even isolate them.

An overwhelming focus of current assistive technology is on tasks like navigation, text reading and general object finding [10] [21] [29]. In the past years attempts have been made in developing social interaction assistants [16] [4] [19]. However, most of these systems take on a technology centric approach i.e., ignoring the user's actual need analysis. An interesting study in identifying the visually impaired user needs in a typical social interaction setting has been performed in [14]. The authors conduct user surveys of 27 visually impaired people by asking questions related to *what is needed*. Among those requirements the few that scored higher includes *(1) I would like to know the names of the people around me. (2) I would like to know how many people there are and where each person is. (3) I would like to know which way each person is facing and when someone looks at me*. A more recent study [1] with 14 visually impaired participants looks at the privacy concerns of using the assistive technology and based on interviews lists features that people with visual impairment would like in a future wearable system. These features also include *counting the number of people nearby and detecting and identify specific face nearby*.

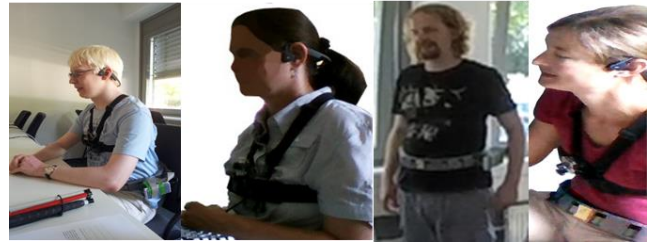

**Figure 1: Our designed system worn by visually impaired participants in a typical social interaction setting**

This provides a good motivation for building systems that can provide these features for the identified needs. Assistive technologies that may provide such non-verbal cues in real time may help boost the confidence and psychological connect in social encounters.

While such a requirement analysis provides a good technology basis it is still needed to actually analyze the impact of such technology's use on the visually impaired. Technology cannot be 100% accurate and thereby may have a counter-intuitive effect on the users causing frustration. It is also needed to see how the users perceive the actual transfer of information via different output interfaces.

The focus of our contribution is to build a system to support social interaction and to analyze its usefulness to the user. To the best of our knowledge this is among the first few studies which brings user in the loop and specifically quantify the impact of using such systems in the typical social interactions. The results make for an important contribution in identifying the actual user response by performing thorough user studies. The study identifies the needed usability aspects of such technology and provides important insights for developing such systems.

In the rest of the paper, after stating the immediate related work we first describe the developed system. We then state the protocol and procedure of the user studies and present the major findings. We conclude after briefly discussing the implications of the studies.

## 2. RELATED WORK

A good survey on wearable assistive devices for the visually impaired can be found in [24]. Most of the existing research has a large focus on building navigational aids for people with visually impairment. An interesting study on assistive technology use in social context has been performed in [22]. The authors concluded that the technology for disabled should address not only usability, function and cost but also the social acceptance. A more recent study [7] on the challenges in the everyday lives of visually impaired people helps understand the need of technology to assist people. The user requirements in social interactions have been identified [14] [1]. Some of the previous works build assistive systems to transfer the identity of the person to the people who are visually impaired in [13] [8] [4] [3] [19]. Most of these only considered transferring identity via speech output. Among these the system presented in [3] and [19] only include a mobile face identification module. Krishna et al. [14] used a pinhole analog CCD camera mounted in a pair of sunglasses. The processing is done in a tablet PC and a text-to-speech converter is used to announce the name of a matched face. The important cues like the direction or the information if someone is looking at the person is not inferred and relayed. The underlying face recognition algorithm is based on a very basic Principal Component Analysis (PCA) and Linear Discriminant Analysis (LDA). The experiments performed are reported in a close-set setting in which the unknown identities are not handled. Astler et al. [3] built a system to infer both identity and expression. They use commercial off-the-shelf systems for face recognition and expression recognition. Their design includes mounting the camera on top of a white cane and the transfer of expression and identity information to the user is just a direct speech output. They also conducts a small user survey with 5 participants to identify the requirements of users and their general feedback of the system, however, no special details or studies are provided on the usability of system's interfaces. The more recent work of Panchanathan et al. [19] also built the system to transfer both identity and expression. They primarily contribute on the algorithm level details for both facial recognition and emotion recognition using active learning schemes. While quantifying the accuracies of the face and emotion recognition on public benchmarks, they do not provide any design on the interfacing part, i.e. how the information should be transferred to the user. Concerning the design of interface, Meers et al. [17] use haptic feedback on fingers to provide information about where the user is looking at by inferring the head orientation. Similarly McDaniel et al. [16] delivers the location and direction of person on a vibrotactile belt. While these systems aim to provide the identified information via speech or haptics, a complete multi-modal system with a thoughtful interface design to answer question like *how*, *when* and *which* information should be transferred to user via multiple interfaces is still missing. In this paper we aim to bridge this gap and provide a real time multi-modal system design along with a new interface design that can successfully transfer the identified visual cues to the users.

## 3. METHODS

We developed a prototype of a new camera-based multi-modal system that assists visually impaired people in social scenarios. Our design is based on the user perceived use cases as has been established from requirement analysis of visually impaired people in social interactions [14] [1] and our interviews with the visually impaired users at the study center for visually impaired at our institute.

Our prototype has two central tasks: 1) it informs the user, on request, of people who are in the camera's field of view: their count, names and position relative to the camera. 2) It informs the user in real-time whenever someone is looking at him, giving information about the name and position of the person relative to the camera. This information about the people in the vicinity is conveyed to the user via an audio-haptic interface.

### 3.1 System Design

The prototype system is comprised of a PC, camera, earphones and a vibrotactile belt. The output devices, i.e., earphones and the belt are wireless, which makes the system more portable, seamless and comfortable for the user. The camera is the only worn-device that is wired. We use a 5 Megapixel Fish eye USB camera with a large field of view of 170 degrees. To avoid covering the ears we use bone-conducting wireless stereo earphones that support panorama stereo for conveying the position of a person relative to the camera. For the haptic output interface we have designed and developed a similar vibrotactile belt as in [16]. It contains 16 equally spaced modules (plastic boxes) with one vibrating motor (cell) inside each. Additional empty modules / boxes can be added as needed, in order to adapt the size of the belt to the participant's waist dimensions. The belt is thus adjustable and includes a battery pack and micro-controller/Bluetooth module to communicate wirelessly with the program. To receive user input (a key press) for when he/she needs information about all the people in the vicinity, we use a keyboard device in this prototype. The user wears the camera (fixated with elastic straps on his chest), earphones and the vibrotactile belt as shown in Figure 1.

### 3.2 Algorithm Design

The system is implemented in C++ under Linux using OpenCV and Dlib libraries. Before proceeding to higher learning tasks such as face detection, recognition etc., it is needed to compensate for the camera movements since it will always be present in a wearable system. We implemented image stabilization using optical flow tracking [15] to compensate for the wearable camera shakes.

**Tell me when someone is looking at me (GAZE):** Participant's gaze direction (where the person is looking at) is important to infer. To this end, in meetings, such as our perceived scenario of social interactions, the eyes of all the participants cannot be tracked reliably from a single wearable camera, and no such system is available in practice. It has been shown [23][18][26][5] that head orientation can be reasonably utilized as an estimate of the gaze when visual focus of attention targets are other meeting participants. We use an implementation of the method in [12] for face detection and facial landmark detection to aid the head orientation estimation. After a face is detected in a frame, a facial landmark localization algorithm is used to search for 68 landmark points on the face such as eye corners, eye centers, lip corners, nose center etc. These 2D point locations can then be used to infer the head orientation by finding the correspondences of these points with points on a fixed 3D model. The landmark detection algorithm is state-of-the-art and can reliably detect landmarks on a face in real time. The underlying algorithm is based on training a random regression forest to accumulate the probabilities of the random facial patches for casting votes to each of the landmark positions on the face. This is captured during the training time by encoding these votes as displacement vectors from each patch center to each of the landmark points. More details about the approach can be seen in Kazemi et al. [12].

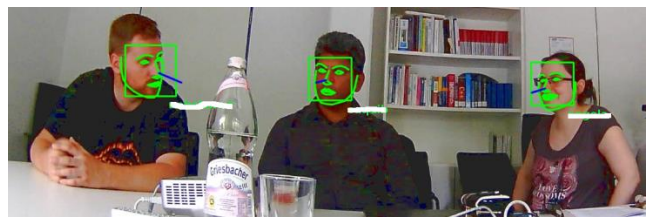

**Figure 2: A typical system output overlaid on a frame. Shows detected faces, recognized person names and the direction where they are looking. (displayed person's names have been anonymized)**

**Table 1: System User interface and functions: ID (tell who is in the room) and Gaze (Tell when someone is looking at me)**

| System variant | Function | Description |
|---|---|---|
| Audio | ID | Speech output: number of people in the room. For each person: Name or Unknown *in stereo sound* |
| Audio | GAZE | Spearcon: [Eye-contact] *in stereo sound* - Speech output: Name or Unknown *in stereo sound* |
| Audio + Haptics | ID | Speech output: number of people in the room. For each person: Name or Unknown *in stereo sound* and *vibration from the direction of the person* |
| Audio + Haptics | GAZE | Spearcon: [Eye-contact] *in stereo sound.* – at the same time a short *vibration from the direction of the person* - Speech output: Name or Unknown *in stereo sound* |

**Tell me who's in the room (ID):** Participant's names (identity) are inferred via face recognition (matching the detected face of the participant with the stored facial images in the database). The face recognition module is implemented by extracting Local Binary Pattern histogram descriptors on the facial fiducial points and training a classifier [28]. The implemented techniques are state-of-the-art and chosen to enable real-time system operation at the frame rate. A typical system output is shown overlaid on a video frame in Figure 2.

### 3.3 Interface Design

A very important aspect of such an assistive system design is to carefully consider how, when and which information should be transferred to users. In the prior attempts of building such systems these considerations are often overlooked and the information is naively transferred to the user via direct speech. In our design, we aim to provide an intelligent interface design that can address the *how, when and which* problem. We set out to define the tasks and functions based on our interviews with the involved focus group.

Table 1 lists use cases for the main functions of the system: ID *Tell me who's in the room* and GAZE *Tell me when someone is looking at me*. For the GAZE function, when someone looks at the user the system speaks out the phrase '*eye contact*' as a spearcon (a very fast spoken text to the point where it is almost indistinguishable) and announces the name using stereo sound. When more than 2 people are looking at the same time towards the user, the system only speaks out the number (e.g., *4 people are looking at you*). This is because we assume that the user is expected to answer a question, or react in another way, and the feedback must be very fast (there is no time to speak out all the names). The user interface was also developed under C++, using the libraries: espeak with the mbrola voices (mbrola-de5) for the speech output, OpenAL (including alut) for stereo panoramic sound and Bluetooth for the connection with the belt. In addition the interface also supports input from the user, in our case a press on the keyboard button, and gives back information about all the people present (in the camera's field of view), while at the same time conveying in real-time the name and position of any person who is currently looking at the user. If a person is unknown to the system i.e., it is not saved in its database, the system speaks out *unknown* instead of the person's name.

### 3.4 System Accuracy

We quantify the system's face identification performance on session recordings from the focus group tests. We also evaluate the head orientation estimation performance on a related public dataset (EGO-HPE) [2] that is collected to evaluate the head pose in social meetings from a wearable camera. The dataset provides a set of egocentric videos with different participants for head pose estimation. Each video is annotated at the frame level for five yaw angles with respect to the participant wearing the camera. The mean head orientation accuracy (in correctly classifying these yaw angles) of our system on the EGO-HPE dataset is 87.3%. This result compares favorably with other state-of-the-art methods published on this dataset. The average face recognition accuracy of our system is found to be 89.13% on the session recordings. These errors of about 10-12% for both head orientation estimation and person identification are indicative of the expected errors in using the current technology.

## 4. The User Study

The purpose of the study is to quantify the usability aspects and the associated cognitive load of such assistive systems on the visually impaired. This will help infer the useful insights not only on developing such systems but also on its applicability in the real world for the target community.

### 4.1 Participants

The user study is conducted with 12 visually impaired participants wearing the system in a typical social interaction setting. The users were chosen among people with visual impairment who could not distinguish the identity of a person (the face traits) as well as exact head orientation, within a distance of one meter and more. The users have an equal gender split i.e., 6 males and 6 females with ages ranging from 22 to 69 years old. The study age group, thus, has an average of 38 and a standard deviation of 16.75 years. 6 from the 12 participants were blind from birth, 2 were late blind and 4 had visual impairment with some remaining sight. All users have used some form of technologies such as smartphone, laptop, tablet or PC on a daily basis. None of them has either a hearing or a tactile impairment (they could all sense the belt's vibration), see Table 2.

### 4.2 Study Procedure

To test the assistive system's impact in a typical social interaction, a discussion is simulated involving the visually impaired participant (the user) and three to five sighted participants. All participants sat around a table and discussed a topic that was agreed upon before the beginning of the study. The topic was moderated by one of the sighted participants, who had the task to keep the conversation going, and also to ask questions from time to time. The participants could address the questions to any of the participant randomly, including the visually impaired participant. To have a realistic meeting scenario, the participants only looked at the person to whom the question was addressed, without mentioning his or her name. Sighted Participants are not particularly told to face directly the user and they sit and discuss naturally as well as free to join and leave the discussion.

Table 2: Users and their backgrounds

| Participant ID | Gender | Age | How often do you use technologies | Your level of experience with technology | Do you play an instrument / sing | Hearing impairment | Tactile impairment | Visual impairment |
|---|---|---|---|---|---|---|---|---|
| 1 | M | 47 | frequent | professional user | yes | no | no | Blind from birth |
| 2 | F | 69 | daily | average user | yes | no | no | Late blind (in the past 10 years or so) |
| 3 | M | 69 | often | average user | no | no | no | Blind from birth |
| 4 | F | 38 | frequent | professional user | no | no | no | Blind from birth |
| 5 | M | 28 | daily | professional user | yes | no | no | Blind from birth |
| 6 | F | 35 | frequent | professional user | no | no | no | Severely visually impaired |
| 7 | F | 47 | frequent | advanced user | yes | no | no | Blind from birth |
| 8 | F | 25 | frequent | professional user | yes | no | no | Visually impaired (10-15% remaining sight) |
| 9 | M | 26 | frequent | advanced user | yes | no | no | Blind from birth |
| 10 | F | 23 | frequent | advanced user | yes | no | no | Severely visually impaired (legally late blind, has a 4% remaining sight and can read. Shortsighted) |
| 11 | M | 28 | daily | professional user | no | no | no | Late blind (from 19 years old) |
| 12 | M | 22 | daily | advanced user | yes | no | no | Severely visually impaired (1.5 - 3% remaining sight) |

During the test, the main task of the visually impaired user was to participate in the discussion, pay attention, make comments and answer questions. The secondary task was to use the system, pay attention and interpret the output.

Each user study took about 2-3 hours to complete. The protocol of the study included a thorough introduction of the study to the user and other participants. The system and its functions are explained to the user and the user is encouraged to try it for a while. The users could adjust the motor strength of the vibrotactile belt, speech and audio in terms of tempo and voice.

In order to fairly evaluate the impact of the system and its different output modalities (audio and haptic interface) the test proceeds in three stages:

1. Testing without the system.
2. Testing with the system worn with audio feedback only.
3. Testing with both audio and haptic feedback.

The order in which the three test stages were carried out was randomized in order to avoid a learning effect.

### 4.3 Questionnaire

Before participating in the tests the user filled in a user background questionnaire (age, visual impairment, level of experience in using technologies).

At the end of each test stage, we assess the cognitive load and the system's usability perception. The user was asked to fill in questionnaires regarding cognitive load (NASA-TLX) [11] and System Usability Scale (SUS) [6]. The SUS was only completed after system worn tests.

At the end of the entire test, the user filled in an overview questionnaire on the usability of interfaces (Audio, Audio + Haptics), functions (ID, GAZE), ease of use and intuitiveness. These concepts are related to the technology acceptance research as laid down in previous studies such as in [9] [25].

## 5. Findings

Here we provide an analysis of the questionnaires and list the main findings. For the statistical comparison the two sided paired t-test is used. The scores reflected below for different tests are measured on a scale from 1 to 5 (1 being low and 5 being high).

## 5.1 Cognitive load

Here we provide the mean NASA-TLX scores (corresponding to the perceived cognitive load) for all participants for the three system variants tests. The test without system resulted in the NASA-TLX score of 3.94. Using System with only audio feedback resulted in a score of 5.58, and system with both audio and vibrotactile feedback scored 7.31. The p-value for the difference between the NASA-TLX scores of the two system variants is 0.0053. This reflects that cognitive load of the users, when participating in a social interaction, among these two system variants is statistically significant. The p-value for the difference between no-system and system (with audio only) is 0.036. The system using both output modalities resulted in a much higher cognitive load. The vibrotactile feedback only complemented the audio feedback while adding one extra channel of information to be processed by the user. However, the vibrotactile information is rated to be more precise by the users.

## 5.2 System Usability Scale (SUS)

The scores for the system usability scale were similar for both the system variants: 71 for system (with audio only) and 68 for the system variant (audio + haptics). According to [20] who analyzed 500 SUS evaluations, "*A SUS score above a 68 would be considered above average and anything below 68 is below average*". As also noted in [27] the mean SUS score of 71 is to be considered as "*good*".

This would place the system in the variant using Audio + vibrotactile feedback as "*average*", while the system variant using only audio is in the "*good*" domain.

However, statistical analysis of the SUS scores shows that there is no significant difference between the two (with p=0.288). One could, therefore, assume that both system variants may be viewed either as "*average*" or "*above average*" at the most from a usability point of view. This is positive for a prototypic assistive system such as the one we used in this study.

## 5.3 Overall Feedback

In the overview questionnaire about the system, all users rated the usefulness of the system in general as well as about its different functions (ID, GAZE). The ease of use and the intuitiveness of the overall system are rated on the higher side with average scores of 4.16 and 4.83 respectively. The overall average score of users' feedback (on usability of interfaces, functions, ease of use and intuitiveness) is 4.31 with an error margin of 0.29. When asked whether they found it better to participate in a meeting using at least one system variant or no system at all, participants answered they preferred to use at least one variant of the system (8 users preferred system with both audio + haptic and 4 users preferred system with audio only).

## 5.4 Observations

During the study, we gather both the comments made by the participants and the observations made by the principal investigator. These are invaluable in assisting further development and provide insights on how such technology can better address the user's need. Participants commented about system's hardware and software. While recognizing that the tested technology is a prototype their observations make for important guidelines for the general technology design in such situations. Many of the users find the hardware *too bulky*. 3 users commented about the camera straps as *too obvious*. Some users preferred camera placement in the shirt's collar or in a name-tag. One user suggested that "*If using straps for the camera put the haptic feedback in these straps*".

With regards to the system functionalities, many users required that they should have the possibility to easily turn off the entire system or only parts of it (e.g., audio) at any time. One user wished the system would provide info only in silent phases of the discussion. Two users perceived the feedback from the system as being "*too much*". For instance, one user reported that

"*While a question was being asked, the system announced that the talking person was looking at me (which meant that I am probably expected to answer). This made me miss one word in the question, and I was unsure if I should answer*".

The aggregated form of GAZE, when more than 2 persons are looking at the same time at the user (e.g., the system output "*4 people are looking at you*") was perceived as useful by the participants who had the occasion to test that function. However, one user (participant ID 9 in Table 2) said

"*The system announces itself in the middle of the conversation, which disturbs a bit;- It has said multiple times in succession: "Eye contact – (name)" once is enough for me, importantly I know, I must react; - Belt would be enough for me (without 3D audio)*"

An observation made by the moderator for this user noted that in the test *without system*: *The respondent has waited for the other participants. If not, then he would answer*. During the test *with system*, *He looked at me when I look in his direction*.

The participants also requested some features that they would like to have in such an assistive system. This includes: *find a friend* (useful in very large group of people); the system should go into a "*quiet mode*" when the user is speaking (either turn itself down completely or turn the audio off); the system should announce when the camera is obscured and has no good view.

Participants also made comments about the vibrotactile feedback. Everyone agreed that in general it is precise and in large meetings it would prove better than audio. Some users preferred the haptic output because for them it was more comfortable and it required less concentration. However, on an average, the belt added more cognitive load on the user. One participant who said he preferred the vibrotactile feedback also said that "*the belt was an extra burden*". One user (participant ID 7 in Table 2) commented that

"*It would be better to use the haptic feedback (without additional audio feedback) in certain situations, such as: when the user speaks or after a while during the meeting, when the user already knows who sits where*".

The reaction time and the vibrotactile feedback of the belt, in general, were regarded as *very good*. We think that it is a very interesting finding based on users' feedback that the haptic modality is more precise than audio feedback. As evident from the users' responses this is important in situations with many people where audio might be less intelligible to the user. Moreover, the vibrotactile actuator can be made much less bulky in a market ready system.

One user (participant ID 4 in Table 2) who really liked the system, commented

"*Camera support pleasant; Voice well; Earphone quite well; - One ingenious idea (sometimes system informs me about a person looking at me, of which I even did not knew he was there); On the smartphone would be quite great, and would use it (without belt,*

*or with a smaller one); It is a relief to have the system"; - So intuitive!*"

## 6. Conclusion

The study showed that the use of assistive technology in social interactions is perceived useful in general. The expected system errors may have a counter-intuitive effect on its users. The study showed that the user can cope with such errors to some extent. The usability analysis revealed some of the missing aspects currently not addressed well in the existing technology e.g., how, when and which information should be transferred to user via multiple interfaces. This study helps understand some of those usability aspects and provide insights for the future design of these systems. Conclusively this study helps connect the technology with the actual visually impaired user's interactions in a typical social interaction scenario. We hope that this effort will bring such technology more close to the market for this important part of society.